\documentclass[fleqn,12pt,draft]{article} 
\usepackage[intlimits,reqno]{amsmath}
\def\oo{\infty}
\def\Z#1{\zeta(#1)}

\def\NuPh{{\it Nucl. Phys. }}
\def\PL{{\it Phys. Lett. }}
\def\PR{{\it Phys. Rev. }}

\def\IJMP{{\it Int. J. Mod. Phys. }}
\def\RMP{{\it Rev. Mod. Phys. }}

\def\gm#1{$g$-$#1$}

\def\e{\epsilon}
\def\ieps{i0}

\def\SYS{{\tt SYS}}
\def\ibp{integration-by-parts }
\def\TITLE{High-precision $\e$-expansions of  three-loop master integrals
contributing to the electron \gm2 in QED }
\def\IDENTIFY{\par (S. Laporta, \TITLE)}
\def\CAPTIONFIG{Topologies of the master integrals. Dotted lines are massless.}

\def\FigureOne
{
\begin{figure}
\begin{center}
   \begin{picture}(390,290)(0,0)
   \thicklines
   \put(035,190){\makebox(0,0)[lb]{
   \put(-020,000){\line(-1,0){8}}
   \put(020,000){\line(+1,0){8}}
   \qbezier(+020.0,+000.0)(+019.9,+008.2)(+014.1,+014.1)
   \qbezier[7](+014.1,+014.1)(+008.2,+019.9)(+000.0,+020.0)
   \qbezier[7](+000.0,+020.0)(-008.2,+019.9)(-014.1,+014.1)
   \qbezier(-014.1,+014.1)(-019.9,+008.2)(-020.0,+000.0)
   \qbezier[7](-020.0,+000.0)(-019.9,-008.2)(-014.1,-014.1)
   \qbezier(-014.1,-014.1)(-008.2,-019.9)(-000.0,-020.0)
   \qbezier(-000.0,-020.0)(+008.2,-019.9)(+014.1,-014.1)
   \qbezier[7](+014.1,-014.1)(+019.9,-008.2)(+020.0,-000.0)
   \put(-014.1,+014.1){\line(+1,-1){28}}
   \qbezier(-014.1,-014.1)(-008,-008)(-002,-002)
   \qbezier(+014.1,+014.1)(+008,+008)(+002,002)
   \qbezier(-002,-002)(-004,+004)(+002,+002)
   \put(-035,-002){\text{$p$}}
   \put(-003,+028){\text{$k_2$}}
   \put(-028,000){\vector(1,0){8}}
   \put(000,020){\vector(1,0){3}}
   \put(-007,-040){\text{\large{$I_1$}}}
   }}
   \put(100,190){\makebox(0,0)[lb]{
   \put(-020,000){\line(-1,0){8}}
   \put(020,000){\line(+1,0){8}}
   \qbezier[15](+020.0,+000.0)(+018.3,+018.3)(+000.0,+020.0)
   \qbezier(+000.0,+020.0)(-018.3,+018.3)(-020.0,+000.0)
   \qbezier(-020.0,+000.0)(-018.3,-018.3)(+000.0,-020.0)
   \qbezier(+000.0,-020.0)(+018.3,-018.3)(+020.0,+000.0)
   \put(000,000){\line(-1,0){20}}
   \put(000,000){\line(0,1){20}}
   \qbezier[7](+020,000)(+010,000)(000,000)
   \put(-007,-040){\text{\large $I_2$ }}
   }}
   \put(165,190){\makebox(0,0)[lb]{
   \put(-020,000){\line(-1,0){8}}
   \put(020,000){\line(+1,0){8}}
   \qbezier(+020.0,+000.0)(+018.3,+018.3)(+000.0,+020.0)
   \qbezier(+000.0,+020.0)(-018.3,+018.3)(-020.0,+000.0)
   \qbezier[15](-020.0,+000.0)(-018.3,-018.3)(+000.0,-020.0)
   \qbezier[15](+000.0,-020.0)(+018.3,-018.3)(+020.0,+000.0)
   \qbezier(000,-020)(+020,000)(000,020)
   \qbezier(000,-020)(-020,000)(000,020)
   \put(-007,-040){\text{\large $I_3$ }}
   }}
   \put(230,190){\makebox(0,0)[lb]{
   \put(-020,000){\line(-1,0){8}}
   \put(020,000){\line(+1,0){8}}
   \qbezier[15](+020.0,+000.0)(+018.3,+018.3)(+000.0,+020.0)
   \qbezier(+000.0,+020.0)(-018.3,+018.3)(-020.0,+000.0)
   \qbezier[15](-020.0,+000.0)(-018.3,-018.3)(+000.0,-020.0)
   \qbezier[15](+000.0,-020.0)(+018.3,-018.3)(+020.0,+000.0)
   \put(000,000){\line(0,1){20}}
   \qbezier[7](-020,000)(-010,000)(000,000)
   \put(000,000){\line(1,0){20}}
   \put(-007,-040){\text{\large $I_4$ }}
   }}
   \put(295,190){\makebox(0,0)[lb]{
   \put(-020,000){\line(-1,0){8}}
   \put(020,000){\line(+1,0){8}}
   \qbezier[15](+020.0,+000.0)(+018.3,+018.3)(+000.0,+020.0)
   \qbezier(+000.0,+020.0)(-018.3,+018.3)(-020.0,+000.0)
   \qbezier(-020.0,+000.0)(-018.3,-018.3)(+000.0,-020.0)
   \qbezier(+000.0,-020.0)(+018.3,-018.3)(+020.0,+000.0)
   \qbezier(-020,000)(-002,+002)(000,+020)
   \qbezier[20](000,-020)(000,000)(000,020)
   \put(-007,-040){\text{\large $I_5$ }}
   }}
   \put(360,190){\makebox(0,0)[lb]{
   \put(-020,000){\line(-1,0){8}}
   \put(020,000){\line(+1,0){8}}
   \qbezier[15](+020.0,+000.0)(+018.3,+018.3)(+000.0,+020.0)
   \qbezier(+000.0,+020.0)(-018.3,+018.3)(-020.0,+000.0)
   \qbezier[15](-020.0,+000.0)(-018.3,-018.3)(+000.0,-020.0)
   \qbezier(+000.0,-020.0)(+018.3,-018.3)(+020.0,+000.0)
   \qbezier(000,-020)(+020,000)(000,020)
   \qbezier[20](000,-020)(-020,000)(000,020)
   \put(-007,-040){\text{\large $I_6$ }}
   }}
   \put(035,100){\makebox(0,0)[lb]{
   \put(-020,000){\line(-1,0){8}}
   \put(020,000){\line(+1,0){8}}
   \qbezier[15](+020.0,+000.0)(+018.3,+018.3)(+000.0,+020.0)
   \qbezier(+000.0,+020.0)(-018.3,+018.3)(-020.0,+000.0)
   \qbezier[15](-020.0,+000.0)(-018.3,-018.3)(+000.0,-020.0)
   \qbezier(+000.0,-020.0)(+018.3,-018.3)(+020.0,+000.0)
   \qbezier[15](-020,000)(-002,+002)(000,+020)
   \put(000,-020){\line(0,1){40}}
   \put(-007,-040){\text{\large $I_7$ }}
   }}
   \put(100,100){\makebox(0,0)[lb]{
   \put(-020,000){\line(-1,0){8}}
   \put(020,000){\line(+1,0){8}}
   \put(000,000){\circle{40}}
   \qbezier(-020,000)(-002,+002)(000,+020)
   \qbezier(+020,000)(+002,-002)(000,+020)
   \put(-007,-040){\text{\large $I_8$ }}
   }}
   \put(165,100){\makebox(0,0)[lb]{
   \put(-020,000){\line(-1,0){8}}
   \put(020,000){\line(+1,0){8}}
   \qbezier[15](+020.0,+000.0)(+018.3,+018.3)(+000.0,+020.0)
   \qbezier(+000.0,+020.0)(-018.3,+018.3)(-020.0,+000.0)
   \qbezier(-020.0,+000.0)(-018.3,-018.3)(+000.0,-020.0)
   \qbezier(+000.0,-020.0)(+018.3,-018.3)(+020.0,+000.0)
   \qbezier(-020,000)(-002,+002)(000,+020)
   \qbezier[15](+020,000)(+002,-002)(000,+020)
   \put(-007,-040){\text{\large $I_9$ }}
   }}
   \put(230,100){\makebox(0,0)[lb]{
   \put(-020,000){\line(-1,0){8}}
   \put(020,000){\line(+1,0){8}}
   \qbezier(+020.0,+000.0)(+018.3,+018.3)(+000.0,+020.0)
   \qbezier(+000.0,+020.0)(-018.3,+018.3)(-020.0,+000.0)
   \qbezier[15](-020.0,+000.0)(-018.3,-018.3)(+000.0,-020.0)
   \qbezier[15](+000.0,-020.0)(+018.3,-018.3)(+020.0,+000.0)
   \qbezier[15](-020,000)(-002,+002)(000,+020)
   \qbezier[15](+020,000)(+002,-002)(000,+020)
   \put(-010,-040){\text{\large $I_{10}$}}
   }}
   \put(295,100){\makebox(0,0)[lb]{
   \put(-020,000){\line(-1,0){8}}
   \put(020,000){\line(+1,0){8}}
   \put(000,000){\circle{40}}
   \qbezier[20](-020,000)(000,+020)(+020,000)
   \qbezier(-020,000)(000,-020)(+020,000)
   \put(-010,-040){\text{\large $I_{12}$}}
   }}
   \put(360,100){\makebox(0,0)[lb]{
   \put(000.3,000){\circle{40}}
   \qbezier(-020,000)(000,+020)(+020,000)
   \qbezier(-020,000)(000,-020)(+020,000)
   \put(-010,-040){\text{\large $I_{13}$}}
   }}
   \put(068,010){\makebox(0,0)[lb]{
   \put(000.3,000){\circle{40}}
   \qbezier[20](-020,000)(000,+020)(+020,000)
   \qbezier[20](-020,000)(000,-020)(+020,000)
   \put(-010,-040){\text{\large $I_{14}$}}
   }}
   \put(133,010){\makebox(0,0)[lb]{
   \put(-028,000){\line(1,0){56}}
   \put(000,000){\circle{40}}
   \qbezier(+027.1,+007.1)(+023.6,+010.6)(020,000)
   \qbezier(+027.1,+007.1)(+030.6,+003.6)(020,000)
   \put(-010,-040){\text{\large $I_{15}$}}
   }}
   \put(198,010){\makebox(0,0)[lb]{
   \put(-028,000){\line(1,0){56}}
   \qbezier[15](+020.0,+000.0)(+018.3,+018.3)(+000.0,+020.0)
   \qbezier[15](+000.0,+020.0)(-018.3,+018.3)(-020.0,+000.0)
   \qbezier[15](-020.0,+000.0)(-018.3,-018.3)(+000.0,-020.0)
   \qbezier[15](+000.0,-020.0)(+018.3,-018.3)(+020.0,+000.0)
   \qbezier(+027.1,+007.1)(+023.6,+010.6)(020,000)
   \qbezier(+027.1,+007.1)(+030.6,+003.6)(020,000)
   \put(-010,-040){\text{\large $I_{16}$}}
   }}
   \put(263,010){\makebox(0,0)[lb]{
   \put(-020,000){\line(-1,0){8}}
   \put(020,000){\line(+1,0){8}}
   \qbezier(+020.0,+000.0)(+018.3,+018.3)(+000.0,+020.0)
   \qbezier(+000.0,+020.0)(-018.3,+018.3)(-020.0,+000.0)
   \qbezier[15](-020.0,+000.0)(-018.3,-018.3)(+000.0,-020.0)
   \qbezier[15](+000.0,-020.0)(+018.3,-018.3)(+020.0,+000.0)
   \qbezier[20](-020,000)(000,+020)(+020,000)
   \qbezier[20](-020,000)(000,-020)(+020,000)
   \put(-010,-040){\text{\large $I_{17}$}}
   }}
   \put(330,010){\makebox(0,0)[lb]{
   \qbezier(-020,000)(-020,+010)(000,000)
   \qbezier(-020,000)(-020,-010)(000,000)
   \qbezier(+010,+017.3)(+001.3,+022.3)(000,000)
   \qbezier(+010,+017.3)(+018.7,+012.3)(000,000)
   \qbezier(+010,-017.3)(+001.3,-022.3)(000,000)
   \qbezier(+010,-017.3)(+018.7,-012.3)(000,000)
   \put(-010,-040){\text{\large $I_{18}$}}   
   }}
   \end{picture}
\end{center}
\phantom{ }\vspace{3truecm}\phantom{ } 
\caption{}
 \label{figureone}
 \end{figure}
}
\newcommand\mytoday{\number\day\space \ifcase\month\or
  January\or February\or March\or April\or May\or June\or
    July\or August\or September\or October\or November\or December\fi
      \space\number\year}

\def\eqref#1{Eq.(\ref{#1})}
\def\eqrefb#1#2{Eqs.(\ref{#1})-(\ref{#2})}

\begin{document}
\title{\vspace{1cm} \TITLE }
\author{S. Laporta\thanks{{E-mail: \tt laporta{\char"40}bo.infn.it}} \\ 
 \hfil \\ {\small \it Dipartimento di Fisica, Universit\`a di Bologna, }
 \hfil \\ {\small \it Via Irnerio 46, I-40126 Bologna, Italy} 
 }
\date{}
\maketitle
\vspace{-7.5cm} \hspace{12.5cm} {\mytoday} \vspace{+7.5cm}
\vspace{1cm}
\begin{abstract}
In this paper we calculate at high-precision the Laurent expansions in $\e=(4-D)/2$ of
the 17 master integrals
which appeared in the analytical calculation of 3-loop QED contribution
to the electron \gm2, using difference and differential equations.
The coefficients of the expansions so obtained 
are in perfect agreement with all the analytical expressions already known.
The values of coefficients not previously known will be used
in the high-precision calculation of the 4-loop QED contribution to the
electron \gm2. 
\end{abstract}

\vspace{1.5cm}
PACS number(s):
\par  12.20.Ds Specific calculations and limits of quantum electrodynamics.
  
\pagenumbering{roman}
\setcounter{page}{0}
\vfill\eject 
\pagenumbering{arabic}
\setcounter{page}{1}

The \gm2 of the electron is probably one of the most precise test of QED.
Over the years, the continuous improvements in the precision of experimental 
determinations
have demanded correspondent improvements of the theoretical predictions.
Let us recall the current status of calculations of QED contribution 
to the electron \gm2.
One- and two-loop contributions are known in closed analytical form 
for a long time.
The calculation of three-loop contribution in closed analytical form 
was completed more recently\cite{3-loop}, after many years of hard work.
Four-loop contribution is known at present only numerically\cite{Kino_last},
with a precision of about 2.5\%, obtained using Monte-Carlo integration methods.
At present this precision is adequate for comparison with current experimental
determinations.  Anyway, there is the need of a new independent 
high-precision calculation, 
in order to cross-check the current numerical value
and to improve considerably its precision, in view of future 
experiments.

A technique which turned out to be useful
in \gm2 calculations is \ibp\cite{Tkachov,Tkachov2}.
The contribution to \gm2 of a graph, 
expressed in the form of a combination of many different integrals
with different powers in the numerator and in the denominator,
is reduced to a linear combination of a small set of ``master integrals''
by using identities obtained by integrating-by-parts in $D$-dimension
space-time.
The master integrals must be calculated analytically or numerically
in the limit $D\to4$ by some method.

In Ref.\cite{3-loop} this technique was applied to the analytical calculation
of the contribution to the electron \gm2 of the last 
family of three-loop graphs still not known analytically,
the so-called triple-cross graphs;
the contributions of all the other three-loop graphs were already obtained
in analytical form by other methods.
The contribution of triple-cross graphs to the \gm2 was reduced
to a combination of 18 master integrals, called $I_1$, $I_2$,{\ldots}, $I_{18}$.
The Laurent expansions in $\e=(4-D)/2$ of the master integrals 
were calculated in analytical form, by
direct calculation (the simpler ones) or by using identities which relate
the coefficients of expansions to values of integrals in 4 dimensions already known
from previous work\cite{light_e,zigzag,ladder,triple_4}.
Subsequently, in Ref.\cite{polon}, we found that the QED contribution of
\emph{all} three-loop
graphs can be reduced to a linear combination of the same master
integrals, which are therefore the only master integrals needed in the three-loop
calculation. We also found that one master integral, $I_{11}$,
is a linear combination of the integrals 
$I_{14}$ and $I_{18}$
so that only 17 master integrals are needed.
The topology of the 17 master integrals is shown in Fig.1.

We plan to use the \ibp method in the calculation of the 4-loop contribution.
We expect about $300\sim400$ master integrals;
an analytical calculation seems to be out of reach, so that alternatively
we consider a  high-precision numerical calculation.
Of course, we begin by calculating the simplest
master integrals, those which factorize in products of master integrals with fewer
loops.
At 3-loop level there are two master integrals, $I_{15}$ and $I_{16}$ (see Fig.1),
which factorize in one 2-loop master integral multiplied by one one-loop tadpole.
At 4-loop, analogously, there are 17 4-loop master integrals which
factorize in the product of one 3-loop master integral $I_j$ and the one-loop tadpole.
The expansion in $\e$ of the one-loop tadpole
contains $1/\e$, so that we need one additional term in the
expansions of the 3-loop master integrals
in order to obtain 4-loop master integrals expanded at the same level
required in the 3-loop calculation.
Unfortunately
the number of terms of the expansions of $I_j$ of
Ref.\cite{3-loop} 
just suffices to obtain the 3-loop result.

Therefore,
for this reason,
in this paper we have calculated at high precision (30 digits)
deeper expansions in $\e$
of the 17 master integrals $I_j$: 
\begin{multline}\label{I1num}
I_1 C^{-1}=
 -.74727427517503872293442889217 
 -5.3830601530185743669980288649 \e
	     \\
 -26.213893908510814642139484925 \e^2
 -105.674804179662879301749904804 \e^3
	     \\
 -385.27172279141197066520099021 \e^4
 -1318.9119175947583164248019535 \e^5
	     \\
 -4337.082974360622762021390292 \e^6
	     +O(\e^7)\ ,
\end{multline}
\begin{multline}
I_2 C^{-1}=
       2.40411380631918857079947632302 \e^{-1}
      -5.33950114034530650865549677321
	     \\
      +25.5217203762232230112382243421 \e
      -73.5271290561422487313093251595 \e^2
	     \\
      +249.044154178834576505276496985 \e^3
      -753.473335692793671251435382464 \e^4
	     \\
      +2346.37248004908075896840609088 \e^5
      -7109.19379761589640023765834979 \e^6
	     +O(\e^7)\ ,
\end{multline}

\begin{multline}
I_3 C^{-1}=
       0.333333333333333333333333333333 \e^{-3}
      +2.33333333333333333333333333333 \e^{-2}
	     \\
      +10.3333333333333333333333333333 \e^{-1}
      +19.7427119912535493212749714262
	     \\
      +88.8774427983382280511485935554 \e
      +95.6785746389971388003677314582 \e^2
	     \\
      +626.528599797078129746832638577 \e^3
      +210.711024315707077431312202546 \e^4
	     \\
      +4305.95682691978475295985357628 \e^5
      -1994.41925210304147109855938156 \e^6
	     +O(\e^7)\ ,
\end{multline}
\begin{multline}
I_4 C^{-1}=
      2.40411380631918857079947632302 \e^{-1}
     -1.88228069596232589686771921947
	     \\
     +24.9738222767976352300663548079 \e
     -40.2430832265114989313433432375 \e^2
	     \\
     +216.453265450196273372723759103 \e^3
     -490.48300355255485916860004085 \e^4
	     \\
     +1877.77603630120111667601080315 \e^5
     -4980.53077825841233320374729215 \e^6
	     +O(\e^7)\ ,
\end{multline}
\begin{multline}
I_5 C^{-1}=
    0.166666666666666666666666666667 \e^{-3}
   +1.5 \e^{-2}
	     \\
   +5.87679853297021379372183633337 \e^{-1}
   +10.2027383130242843587260466742 
	     \\
   +49.4062940751534383431822667245 \e
   +38.7587897454906373636992141483 \e^2
	     \\
   +355.54075314113673706462779833 \e^3
   -12.8313984708255340018628252229 \e^4 
	     \\
   +2544.39228074929353310098747381 \e^5
   -2170.06912961044909561870930877 \e^6
	     +O(\e^7)\ ,
\end{multline}
\begin{multline}
I_6 C^{-1}=
           0.333333333333333333333333333333 \e^{-3}
	  +2.33333333333333333333333333333 \e^{-2}
	     \\
	  +10.3333333333333333333333333333 \e^{-1}
	  +29.7659868661137435310832928686
	     \\
	  +102.438098206123109827196396063 \e
	  +229.885864095858993939826105646 \e^2
	     \\
	  +762.395106594671638134114238656 \e^3
	  +1423.01797639907075045462241083 \e^4
	     \\
	  +5052.27961041516425974606724499 \e^5
	  +7498.83181107570646929366512712 \e^6
	     +O(\e^7)\ ,
\end{multline}
\begin{multline}
I_7C^{-1}=
                0.166666666666666666666666666667 \e^{-3}
	       +1.5 \e^{-2}
	     \\
	       +5.87679853297021379372183633337 \e^{-1}
	       +18.0613667204622021855023607236
	     \\
	       +55.4913080622726427983915540249 \e
	       +134.432560982177142077741730996 \e^2
	     \\
	       +394.304047315939025666187053151 \e^3
	       +824.935548393665344225617774097 \e^4
	     \\
	       +2505.10730543388776893266213278 \e^5
	       +4469.58551067612956096036609952 \e^6
	     +O(\e^7)\ ,
\end{multline}
\begin{multline}
I_8C^{-1}=
                  -  \e^{-3}
		  -5.33333333333333333333333333333 \e^{-2}
		  -16 \e^{-1}
	     \\
		  -43.9148312632524344127591663433
		  -154.918028662961860140720725441 \e
	     \\
		  -374.094185333355168755155126113 \e^2
		  -1436.67271253544219472250003758 \e^3
	     \\
		  -3281.94043631920573984798601407 \e^4
		  -13100.0652782910379543889228475 \e^5
	     \\
		  -29197.1252330882109400594603379 \e^6
	     +O(\e^7)\ ,
\end{multline}
\begin{multline}
I_9C^{-1}=
          -0.666666666666666666666666666667 \e^{-3}
	  -3.33333333333333333333333333333 \e^{-2}
	     \\
	  -11.956534800363119539611497 \e^{-1}
	  -44.5995196208454844578584038609
	     \\
	  -125.546733075809915579874065329 \e
	  -442.029796112593386030203112952 \e^2
	     \\
	  -1165.46537028123770794611984172 \e^3
	  -4078.5492471997365803005187143 \e^4
	     \\
	  -10533.3774099326183047654759014 \e^5
	  -36961.4052223824089107170672131 \e^6
	     +O(\e^7)\ ,
\end{multline}
\begin{multline}
I_{10}C^{-1}=
          -0.333333333333333333333333333333 \e^{-3}
	  -1.66666666666666666666666666667 \e^{-2}
	     \\
	  -10.5797362673929057458896606666 \e^{-1}
	  -30.1135700965916539174115430661
	     \\
	  -136.273953266205829743326933622 \e
	  -333.735442456113204996932778211 \e^2
	     \\
	  -1376.89949509222838398713383632 \e^3
	  -3194.567378327629536037718821 \e^4
	     \\
	  -12862.3364927028759414511191045 \e^5
	  -29234.6901535778480407834346202 \e^6
	     +O(\e^7)\ ,
\end{multline}
\begin{multline}
I_{12}C^{-1}=
             \e^{-3}
	    +3.5 \e^{-2}
	    +7.02777777777777777777777777778 \e^{-1}
	     \\
	    +11.5787037037037037037037037037
	    -24.3219708028082785734403310608 \e
	     \\
	    -121.961009709497309747947635538 \e^2
	    -819.989946014746295714157758046 \e^3
	     \\
	    -2404.86316062636537480556898164 \e^4
	    -10421.8794385530161287164813459 \e^5
	     \\
	    -27852.0785333053790606284709671 \e^6
	     +O(\e^7)\ ,
\end{multline}
\begin{multline}
I_{13}C^{-1}= 2 \e^{-3}
            +7.66666666666666666666666666667 \e^{-2}
	    +17.5 \e^{-1}
	     \\
	    +22.9166666666666666666666666667
	    +21.2517910512915199882568913631 \e
	     \\
	    -184.230005105298483421085631284 \e^2
	    -661.110586153353833254708841052 \e^3
	     \\
	    -3685.05477938169880569422657784 \e^4
	    -10050.975403938380587540545542 \e^5
	     \\
	    -42319.9745464132618093606637193 \e^6
	     +O(\e^7)\ ,
\end{multline}
\begin{multline}
I_{14}C^{-1}= 0.333333333333333333333333333333 \e^{-3}
            +1.16666666666666666666666666667 \e^{-2}
	     \\
	    +2.08333333333333333333333333333 \e^{-1}
	    +2.99715174175891809439930176403
	     \\
	    -21.7478477083774449677944881844 \e
	    -84.9528451829928094897709758796 \e^2
	     \\
	    -520.302648810090620535897692307 \e^3
	    -1504.0311391147000649985117616 \e^4
	     \\
	    -6482.74483328181932000540551865 \e^5
	    -17315.8330540574013027834356625 \e^6
	     +O(\e^7)\ ,
\end{multline}
\begin{multline}
I_{15}C^{-1}= 1.5 \e^{-3}
            +5.75 \e^{-2}
	    +13.125 \e^{-1}
	    +30.3469725347858114917793213332
	     \\
	    +59.9061795885052938207577452894 \e
	    +128.784074758885330304917817688 \e^2
	     \\
	    +247.030559404244292769216590768 \e^3
	    +522.536074129783764428998974715 \e^4
	     \\
	    +995.526049616742468296298492052 \e^5
	    +2097.54543056218400478046531901 \e^6
	     +O(\e^7)\ ,
\end{multline}
\begin{multline}
I_{16}C^{-1}= 0.5 \e^{-3}
            +1.75 \e^{-2}
	    +6.41486813369645287294483033329 \e^{-1}
	     \\
	    +16.0102660805759621969058588126
	    +42.0560659585935725800136421336 \e
	     \\
	    +91.5580001461564241262070311808 \e^2
	    +211.940358797491166303979686218 \e^3
	     \\
	    +432.923773655861990482605036606 \e^4
	    +946.511285468009699310821028163 \e^5
	     \\
	    +1874.61717426917014415736822215 \e^6
	     +O(\e^7)\ ,
\end{multline}
\begin{multline}
I_{17}C^{-1}=  -0.166666666666666666666666666667 \e^{-2}
             -0.972222222222222222222222222222 \e^{-1}
	     \\
	     -5.87783109665941583590779329626
	     -23.4899848806977618982607310895 \e
	     \\
	     -97.5806595141990146514480659561 \e^2
	     -330.869679035399716046952939963 \e^3
	     \\
	     -1181.19322772402976499602304607 \e^4
	     -3694.74722374791467371175473492 \e^5
	     \\
	     -12281.2292376715157620002229398 \e^6
	     +O(\e^7)\ ,
\end{multline}

\begin{multline}\label{I18num}
I_{18}C^{-1}= -\e^{-3} -3 \e^{-2} -6 \e^{-1} -10 -15 \e -21 \e^2 -28 \e^3 -36 \e^4
-45 \e^5 -55 \e^6 +O(\e^7) \ .\qquad\qquad
\end{multline}
The integrals $I_j$ are defined as\cite{3-loop}
\begin{equation*}
I_j=
  \left(\dfrac{-i}{\pi^{D-2}}\right)^3
  \int{d^D k_1 d^D k_2 d^D k_3}\ \dfrac{P_j}{Q_j} \ ,
  \qquad P_1=p\cdot k_2\ , \quad P_j=1 \text{ if } j\ge2\ , 
\end{equation*}
$Q_j$ contains the product of
the denominators of the corresponding $j$-th graph of Fig.1.
The normalization factor is $C=\left(\pi^{\e}\Gamma(1+\e)\right)^3$.
The number of terms of the expansions of \eqrefb{I1num}{I18num} 
suffices for the use in
\gm2 calculations at four and even more loops (actually,
we have calculated much deeper expansions, not completely listed here for lack of space).

\eqrefb{I1num}{I18num} are to be compared with the corresponding analytical
results of Ref.\cite{3-loop}.
Due to an unfortunate misprint (see \cite{polon}),
in Ref.\cite{3-loop} the terms containing
the constants $C_1$, $C_2$ are missing in the {\it r.h.s.} of the integrals
$I_2$, $I_3$, $I_4$, $I_5$, $I_6$, $I_7$ and $I_{11}$; all the results
which follow are however correct, as $C_1$, $C_2$ cancel out systematically
in the final results of \cite{3-loop}.
For the ease of the reader we list here the correct expressions
of the integrals $I_2$, $I_3$, {\ldots}, $I_7$
(already appeared in \cite{polon});
the remaining integrals are listed in Ref.\cite{3-loop}:
\begin{multline}\label{I2anal}
I_2 C^{-1}= 2 \frac{\Z3 }{\e}
       - \frac{13}{90} \pi^4 - \frac{1}{3} \pi^2 + 10 \Z3 
   + \e \biggl( \frac{385}{2} \Z5 
 -\frac{85}{6} \pi^2 \Z3  
 \\
   -\frac{7}{15} \pi^4 
                 -{82 \Z3 } -{4 \pi^2 {\ln 2}} 
   +{16} \pi^2 -2 C_1
 +6 C_2 \biggr) +O(\e^2)  \ ,
\end{multline}
\begin{multline}
I_3 C^{-1}=
    \frac{1}{3\e^3} +  \frac{7}{3\e^2} +  \frac{31}{3\e} 
      - \frac{2}{15} \pi^4 - \frac{4}{3} \Z3 + \frac{103}{3} 
   + \e \biggl( 95\Z5 
     - \frac{25}{3}\pi^2\Z3 - \frac{1}{15}\pi^4 
     \\
              - \frac{184}{3} \Z3 - 8 \pi^2{\ln 2} 
   + \frac{44}{3}\pi^2 
              + \frac{235}{3} +4 C_2 \biggr) 
    +O(\e^2)     \ ,
\end{multline}
\begin{multline}
I_4 C^{-1}=  
    2 \frac{\Z3}{\e} - \frac{7}{90} \pi^4 + 2 \Z3 + \frac{1}{3} \pi^2  
   + \e \biggl( \frac{385}{2} \Z5
 -\frac{85}{6} \pi^2 \Z3 
   -\frac{7}{15} \pi^4 
     \\
                 -{82 \Z3 }
    -{4 \pi^2 {\ln 2}} +{16} \pi^2 -2 C_1
 +4 C_2
                   \biggr) +O(\e^2)  \ , 
\end{multline}
\begin{multline}
I_5 C^{-1}=  
        \frac{1}{6\e^3} + \frac{3}{2\e^2}
       + \frac{1}{\e} \biggl(- \frac{1}{3} \pi^2 + \frac{55}{6} \biggr)
       - \frac{4}{45} \pi^4 - \frac{14}{3} \Z3
 - \frac{7}{3} \pi^2 
                                             + \frac{95}{2}  
     \\
       + \e   \biggl( - \frac{2}{9} \pi^4 - 44 \Z3 - \frac{29}{3} \pi^2 
                         + \frac{1351}{6}
  +2 C_1 \biggr) +O(\e^2)  \ ,
\end{multline}
\begin{multline}
I_6 C^{-1}=
       \frac{1}{3\e^3} 
       + \frac{7}{3\e^2} 
       + \frac{31}{3\e} 
       - \frac{4}{45} \pi^4  + \frac{2}{3} \Z3 + \frac{1}{3} \pi^2 + \frac{103}{3}
       + \e   \biggl(
       \frac{45}{2} \Z5 - \frac{7}{2} \pi^2\Z3 
+ \frac{11}{45} \pi^4
     \\
+ \frac{14}{3} \Z3  - 4 \pi^2{\ln 2} 
    + \frac{14}{3} \pi^2 
    + \frac{235}{3} 
    +2C_1 \biggr) +O(\e^2)  \ ,   
\end{multline}
\begin{multline}\label{I7anal}
I_7 C^{-1}=
        \frac{1}{6\e^3}
       + \frac{3}{2\e^2}
       + \frac{1}{\e} \biggl( - \frac{1}{3} \pi^2 + \frac{55}{6}  \biggr)
       - \frac{1}{15} \pi^4 - \frac{8}{3} \Z3
 - 2 \pi^2 
     + \frac{95}{2}
      + \e   \biggl(
     \frac{45}{2} \Z5 
     - \frac{17}{6} \pi^2\Z3 
     \\
   - \frac{7}{9} \pi^4
   - 50 \Z3 
   - 4 \pi^2{\ln 2} + \frac{1}{3} \pi^2 
   +\frac{1351}{6} 
     +2C_2\biggr) 
   +O(\e^2)    \ ,  
\end{multline}
where $\zeta(p)=\sum_{n=1}^\oo 1/n^p$, 
$a_4=\sum_{n=1}^\oo \tfrac{1}{2^n n^4}$, and
$C_1$, $C_2$ are the coefficients of the $\e$ term in the expansion 
of the integrals
\begin{eqnarray}
 M_1 &=& 
 \left(\dfrac{-i}{\pi^{D-2}}\right)^3
 \int\dfrac{d^D k_1 d^D k_2 d^D k_3\ (p.k_1)}{D_1 D_3 D_4 D_5 D_6 D_7 D_8 } 
   = C\left( - \frac{2}{45}\pi^4  +\Z3 + \e C_1 +O(\e^2) \right) , 
                                                               \nonumber \\\cr
 M_2 &=& 
  \left(\dfrac{-i}{\pi^{D-2}}\right)^3
  \int\dfrac{d^D k_1 d^D k_2 d^D k_3\ (p.k_3)}{D_2 D_3 D_4 D_5 D_6 D_7 D_8 } 
   =C\left( - \frac{1}{30}\pi^4 - \frac{1}{3}\pi^2 + 4 \Z3 +\e C_2 
                                           +O(\e^2) \right)\ ,\nonumber \\\cr
\nonumber
\end{eqnarray}
      $D_1 = (p-k_1)^2+1-\ieps$,          $D_2 = (p-k_1-k_2)^2+1-\ieps$,
      $D_3 = (p-k_1-k_2-k_3)^2+1-\ieps$,  $D_4 = (p-k_2-k_3)^2+1-\ieps$,
      $D_5 = (p-k_3)^2+1-\ieps$,          $D_6 = k_1^2-\ieps$,
      $D_7 = k_2^2-\ieps$,                $D_8 = k_3^2-\ieps$,
      $p^2=-1$. 

The analytical values of $C_2$ and $C_1$ were calculated, respectively,
in Ref.\cite{polon} and Ref.\cite{Ritbergen}
\begin{equation*}
 C_2 = - \frac{173}{4}\Z5 + \frac{53}{12}\pi^2\Z3  - \frac{2}{15}\pi^4
          + 18 \Z3 + 2 \pi^2 \ln 2  - 3 \pi^2  \ ,
\end{equation*}	  
\begin{equation*}
C_1 = - \frac{49}{4}\Z5 + \frac{25}{12}\pi^2\Z3  - \frac{49}{180}\pi^4
          + \Z3 + 2 \pi^2 \ln 2  - \frac{\pi^2 }{3} \ .
\end{equation*}	  

\eqrefb{I1num}{I18num}
agree perfectly with
the analytical expressions
of \eqrefb{I2anal}{I7anal} and of Refs.\cite{3-loop,polon}, and with the deeper
expansions of
$I_{10}$\cite{Ritbergen},
$I_{13}$, $I_{15}$\cite{Broadhurst}, $I_{16}$ and $I_{17}$.

Now we sketch the method used for obtaining \eqrefb{I1num}{I18num}.
We have used in this calculation the new method of calculation of
master integrals based on solution of difference equations in exponents
developed and described in detail in Refs.\cite{Lapdif1,Lapdif2}.
This method consists in the construction and numerical solution 
of systems of difference equations
between the master integrals $I_j$
(with polynomials in $D$ and $n$ as coefficients),
seen as functions of the exponent $n$ of one denominator.
The difference equation
for a given master integral
contains in the {\it r.h.s.} only master integrals with less denominators,
which are simpler; therefore
the equations of the system can be resolved one at once, beginning with
that corresponding to the simplest master integral and ending with that
corresponding to the most complex.
Suitably boundary conditions at $n\to\oo$ must be provided for
a proper solution of the difference equations. This implies
the calculation of integrals with one loop less,
which can be obtained by solving other systems of difference equations.
The amount of calculations needed to work out and solve the systems of
difference equations is rather high, so that the practical application of
the method relies on the use of an automatic tool, the program $\SYS$
described in Ref.\cite{Lapdif1}.

Because of some actual limitations of the program $\SYS$,
we have found it convenient to give a mass $\lambda\not=0$
to the photon lines, so that the
master integrals become  functions $I_j(\lambda)$.
Then, we have used the above described method to calculate the value of master integrals
$I_j(1)$
(actually, these values were already calculated using this method
in Refs.\cite{Lapdif1-9}; here we have repeated the same calculation
increasing the numerical precision and the order of expansions in $\e$). 

Subsequently, always  by means of the program $\SYS$\cite{Lapdif1-10},
we have built a system 
of differential equations in $\lambda$ for $I_j(\lambda)$ 
(the approach based on the use of differential
equations in masses was introduced in Ref.\cite{difmas}),
and we have integrated it, using as initial conditions the values 
(and, when needed, the derivatives w.r.t. $\lambda$) at $\lambda=1$,
obtaining the values $I_j(0)$, that is, \eqrefb{I1num}{I18num}.

We note that the systems of difference and
differential equations for $\lambda\not=0$ are much more complicated 
(more master integrals, higher order equations, higher degree coefficients) 
than the system of difference equations for the original master 
integrals with zero masses $I_j(0)$, so that the solution of the latter
would be preferable for calculating 4-loop master integrals which do not
factorize.
We plan to overcome this limitation in near future.


\vfill\eject 
\pagenumbering{roman}
\setcounter{page}{1}
\phantom{.}\vspace{5cm}
\section*{Figure Captions}
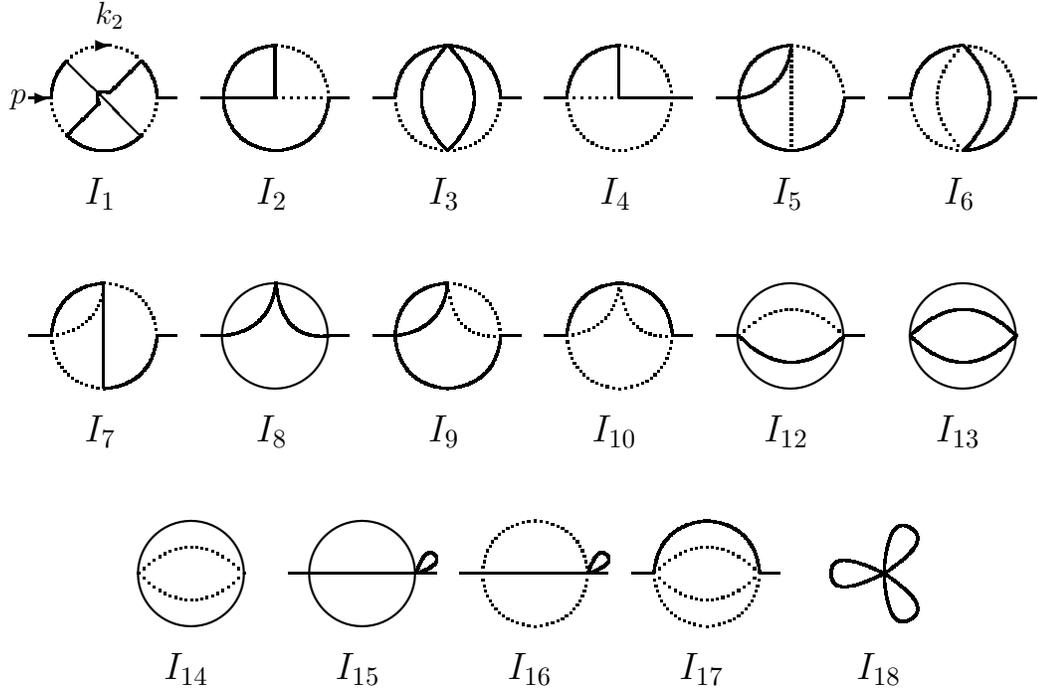
\par\noindent Figure 1: \CAPTIONFIG
\phantom{.}\vspace{12cm}\IDENTIFY
\phantom{.}
\vfill\eject 
\phantom{.}\vspace{1cm}
\FigureOne
\phantom{.}\vspace{1cm}\IDENTIFY
\end{document}